\documentclass[floats,floatfix,showpacs,amssymb,prd,twocolumn,superscriptaddress,nofootinbib,longbibliography]{revtex4-1}

\usepackage{anyfontsize}
\usepackage{amssymb}
\usepackage{amsmath}
\usepackage{amsfonts}
\usepackage{amsbsy}
\usepackage[utf8]{inputenc} 
\usepackage{newtxtext}
\usepackage{newtxmath}
\usepackage{lipsum,graphicx,afterpage}
\usepackage{tensor}
\usepackage{mathrsfs}

\usepackage{bm}

\usepackage[utf8]{inputenc}

\usepackage{graphicx}
\usepackage{epsfig}
\usepackage{epstopdf}

\usepackage[usenames,dvipsnames]{xcolor}

\usepackage{verbatim,mathtools,needspace,enumitem,etoolbox}

\definecolor{linkcolor}{rgb}{0.0,0.3,0.5}

\usepackage[unicode, colorlinks=true, linkcolor=linkcolor, citecolor=linkcolor, filecolor=linkcolor,urlcolor=linkcolor, pdfusetitle]{hyperref}

\usepackage{epstopdf}

\usepackage{bm}
\usepackage{dcolumn}

\usepackage{longtable}
\usepackage{enumerate}
 \usepackage[normalem]{ulem}

\hypersetup{colorlinks=true,
            citecolor=blue,
            linkcolor=blue,
            urlcolor=blue}
\usepackage[T1]{fontenc}
\usepackage{subfigure}
\usepackage{amsmath}
\usepackage{float}
\usepackage{mathrsfs}
\usepackage[dvipsnames]{xcolor}
\usepackage{soul}

\definecolor{darkgreen}{RGB}{1,212,57}

\definecolor{darkblue}{RGB}{50,205,50}

\newcommand{\be}{\begin{equation}}
\newcommand{\ee}{\end{equation}}

\setcitestyle{numbers,square}

\setcitestyle{numbers,square}

\begin{document}
\title{Tidal forces in dirty black hole spacetimes}

\author{Haroldo C. D. Lima Junior}
\email{haroldo.ufpa@gmail.com} 
\affiliation{Faculdade de F\'{\i}sica,
Universidade Federal do Par\'a, 66075-110, Bel\'em, PA, Brasil }

\author{Mateus M. Corr\^ea}
\email{mateus.correa@icen.ufpa.br} 
\affiliation{Faculdade de F\'{\i}sica,
Universidade Federal do Par\'a, 66075-110, Bel\'em, PA, Brasil }

\author{Caio F.~B.~Macedo}
\email{caiomacedo@ufpa.br} 
\affiliation{Faculdade de Física, Campus Salin\'opolis, Universidade Federal do Par\'a, 68721-000, 	Salin\'opolis, Par\'a, Brazil}

\author{Lu\'{\i}s C. B. Crispino}
\email{crispino@ufpa.br} 
\affiliation{Faculdade de F\'{\i}sica,
Universidade Federal do Par\'a, 66075-110, Bel\'em, PA, Brasil }

\begin{abstract}
Black holes can be inserted in very rich astrophysical environments, such as accretion disks. Although isolated black holes are simple objects in general relativity, their accretion disks may significantly enrich the field configurations of their surroundings.
Alternative theories of gravity can lead to novel black hole solutions, which can be represented by small deviations in the metric due to an effective stress-energy tensor. 
Among the key aspects of the interaction of black holes with their surroundings, stand tidal forces phenomena.
We study the tidal forces of spherically symmetric black holes in the presence of effective matter fields, dubbed as {\it dirty black holes}. These effective fields can generically represent usual or exotic matter associated to a variety of gravity theories.
We show that this {\it dirtiness} leads to characteristic imprints in the tidal forces, which are absent in the case of a black hole surrounded by vacuum. We apply our results to particular cases, such as 
black holes coupled to linear and nonlinear electrodynamics theories and 
a Schwarzschild black hole surrounded by a spherical shell.
\end{abstract}

\maketitle

\section{Introduction}
\label{sec:int}

General relativity (GR) introduced new objects to the scientific knowledge, such as black holes (BHs). In the eletrovacuum, isolated BHs may be described by solely three parameters: mass, charge and angular momentum \cite{Uniq}. Although astrophysical BHs are expected to be described by only two of these parameters --- mass and angular momentum --- they may harbor very dynamical environments, such as accretion disks and electromagnetic fields~\cite{frank}. Moreover, in alternative theories of gravity BHs are associated to slight deviations from their {\it bald} counterpart, which can leave distinctive imprints in their phenomenology~\cite{Berti:2015itd}. The study of how these deviations can manifest themselves in the electromagnetic and gravitational waves observations constitutes a current intense research subject in the literature~\cite{Barausse2014,Berti:2019wnn,Ferreira:2017pth}.

In the past years, the strong regime of gravity has finally started to be tested, and the results are in good agreement with the GR predictions \cite{EHT1,Abbot2016}.
The shadow recently imaged by the international collaboration {\it Event Horizon Telescope} (EHT) is associated to a BH with an accretion disk, which casts an asymmetric bright emission configuration~\cite{EHT1}. A similar image is expected for other astrophysical BHs hosted in galactic centers. The presence of matter and fields in the neighborhood of BHs is of astrophysical importance, and must be taken into account to identify BH candidates. Tidal effects on objects near the event horizon have been proposed to be the cause of flares \cite{cadez2009}, as optical~\cite{TDE1}, ultraviolet~\cite{TDE2} and X-ray radiation~\cite{TDE3}, during tidal disruption events. 

It is well-known that tidal forces in BH spacetimes cause the stretching and compressing of extensive objects. In the case of a Schwarzschild BH, an infalling body is stretched along the radial direction and compressed along the angular directions \cite{Pirani,Manasse,dinverno,Hobson}. For a Reissner-Nordstr\"om (RN) BH the tidal forces can vanish at some positions and change from stretching to compressing, and the other way around, depending on which component of the tidal force one is analyzing \cite{crispino2016}. These features are also manifest in other 
spacetimes, such as
regular~\cite{msharif2018,Junior:2020IJMPD}, Kiselev with radiation or dust~\cite{Umair2017}, Kottler geometry~\cite{kottler2021} and Kerr BHs~\cite{Junior:2020yxg,R2,R5,R6} . For some of these spacetimes, the radial coordinate where the tidal force vanishes can be located outside the event horizon, leading, in principle, to observable features. The change in the sign of tidal forces can have interesting astrophysical consequences, such as modifications of the tidal disruption radius, also known as Roche radius~\cite{1974ApJ...191..577A,1972ApJ...175..L155,1985RAstrSoc...212..57}.

Matt Visser introduced a way to study BHs surrounded by some classical matter fields, dubbed as {\it dirty BHs} \cite{Visser1992}. 
For instance, the
RN BH may be regarded as a dirty BH, with the static electric field surrounding the BH being associated to the {\it dirtiness}. 
Hence, the difference between the tidal forces in Schwarzschild and RN BHs can be interpreted as due to this dirtiness. 
The influence of such dirtiness in BH physics has been investigated for the Hawking temperature \cite{Visser1992}, quasinormal modes \cite{Leung1999,Visser2004,VisserSqueezed2004}, absorption \cite{Caio2016}, scattering \cite{Caio2019}, and gravitational wave astronomy \cite{Barausse2014}.
Following this idea,  
we investigate tidal forces for different kinds of dirty BHs. 
We assume that the dirtiness around BHs may be effectively described by the stress-energy tensor of an anisotropic fluid, and
study the influence of the anisotropy and the energy density on the radial and angular tidal forces. 
For instance, we show that, depending on the energy density and pressure profiles of the dirtiness, tidal forces can vanish at some points, what may influence, for instance, the Roche radius.

The remaining of this paper is organized as follows. In Sec.~\ref{sec2} we start from a generic spherically symmetric spacetime, and obtain the relations between the metric functions and the properties of the dirtiness surrounding the BH. In Sec.~\ref{sec2.1} we represent the matter-energy components by a generic anisotropic fluid compatible with spherically symmetric spacetimes. In Sec.~\ref{sec3} we study the radial geodesics for such configuration. In Sec.~\ref{sec4} we introduce the geodesic deviation equation, identify the tetrad basis attached to an observer radially infalling toward the BH, and obtain the tidal tensor for radially infalling bodies. In Sec.~\ref{sec5} we analyze some examples, and investigate the tidal forces in each case. We present our final remarks in Sec.~\ref{sec6}. Throughout this paper, we use geometrical units ($c = G = 1$) and the metric signature ($-$, $+$, $+$, $+$).

\section{Dirty black hole spacetimes}\label{sec2}

\subsection{Spacetime}
\label{sec2.1}

We consider a generic spherically symmetric spacetime, described by the following line element: 
\be
\label{lineel}ds^2=-f(r)\,dt^2+\frac{dr^2}{h(r)}+r^2\,d\Omega^2,
\ee
where we have used Schwarzschild-like coordinates, with $d\Omega^2$ being the 2-sphere solid angle element. We assume $f(r)$ and $h(r)$ to be two strictly positive functions outside the event horizon, satisfying asymptotically flatness conditions, with $f(r)\rightarrow 1$ and $h(r)\rightarrow 1$, as $r\rightarrow \infty$. We assume the existence of an event horizon, obeying
\begin{align}
	\label{horizon_cond1}&h(r_+)=0,\\
	\label{horizon_cond2}&h'(r_+)\geq 0.
\end{align}
We refer to spacetimes in which the equality in Eq.~\eqref{horizon_cond2} holds as \textit{extreme dirty BHs}. 
We shall write the function $h(r)$ as
\be
h(r)\equiv 1-\frac{2b(r)}{r},
\ee
where $b(r)$ is the so-called Misner-Sharp energy within a sphere of radius $r$ \cite{2009PhyRevD...80..104016}. At the radial coordinate of the event horizon, we have 
\begin{align}
	&b(r_{+})=\dfrac{r_{+}}{2},\\
	&2b'(r_+)\leq 1.
\end{align}
We assume that the line element (\ref{lineel}) is a solution of a gravity theory with equations of motion which can be written in the form
\be
\label{Einstein_eqs}G^{\mu}\,_{\nu}=8\,\pi T^{\mu}\,_{\nu},
\ee
where $G^{\mu}\,_{\nu}$ are the components of the Einstein tensor~\cite{Wald::1984} and $T^{\mu}\,_{\nu}$ is an effective stress-energy tensor.
Note that in this scenario the stress-energy tensor $T_{\mu\nu}$ may collectively account for matter sources and possible modified gravity terms.\footnote{We emphasize that, although not all extended theories of gravity can be described by an effective stress-energy tensor, we assume that this is true at least perturbatively, in the sense that we recover GR as a particular case~\cite{Clifton:2011jh}.	} 
We also assume, for simplicity, that the dirtiness can be described by an effective anisotropic fluid~\cite{Boonserm:2015aqa}, with the stress-energy tensor being
\begin{equation}
	{T^{\mu}}_{\nu}={\rm diag}(-\rho,p,q,q),
\end{equation} 
where $\rho$ is the energy density, $p$ the radial pressure and $q$ the tangential pressure.  We note, in passing, that scalar and electromagnetic fields sourcing gravity can be represented by anisotropic fluids~\cite{Boonserm:2015aqa}.  
Due to the spherical symmetry, the stress-energy tensor components ${T^{\mu}}_{\nu}$ are functions only of the radial coordinate $r$, and the angular stresses are equal (i.e., ${T^{\theta}}_{\theta}={T^{\phi}}_{\phi}$).

By computing $G^{\mu}\,_{\nu}$ for the line element~\eqref{lineel} and inserting it into the field equations~\eqref{Einstein_eqs}, we find that
\begin{align}
	\label{Gtt}&2\,b'= 8\,\pi\,r^2\,\rho,
\end{align}
\begin{align}
	&{\left(\frac{r\,f'}{f}-\frac{2\,b}{r-2\,b}\right)\left(1-\frac{2\,b}{r}\right)= 8\,\pi\,r^2\,p },
\end{align}
\begin{align}
	\label{Gthth}\frac{b}{r}-b'-&\frac{r\,\left(r-2\,b\right)\,f'^2}{4\,f^2}+ \nonumber
	\\+&\frac{r\,f''\,\left(r-2\,b\right)-f'\left[b+r\,\left(b'-1\right)\right]}{2\,f}=8\,\pi\,r^2\,q,
\end{align}
where the primes denote differentiation with respect to the radial coordinate $r$. From the conservation of the stress-energy tensor, i.e. $\nabla_\mu\,T^{\mu\nu}=0$, we obtain
\be
\label{divT}r\,f'\,(\rho+p)+2\,f\left(2\,\sigma+r\,p'\right)=0,
\ee
where $\sigma\equiv p-q$ gives a measure of the anisotropy in the stress-energy tensor. It is useful to write  $f'$, $f''$, $b'$ and $p'$ in terms of ($r$, $b$, $f$, $\rho$, $p$, $q$). Using Eqs.~\eqref{Gtt}-\eqref{divT}, we find that
\begin{align}
	\label{mu'}&b'=4\,\pi\,r^2\,\rho,
\end{align}
\begin{align}
	\label{f'}&f'=\frac{2\,f\left(4\,\pi\,r^3\,p+b\right)}{r\,\left(r-2\,b\right)},
\end{align}
\begin{align}
	\label{p'}p'=-\frac{1}{r\left(r-2\,b\right)}\left[4\,\pi\,r^3\,p^2-2\,q\left(r-2\,b\right)+b\,\rho \right. \nonumber \\
	+ \left. p\left(2\,r+4\,\pi\,r^3\,\rho-3\,b\right) \right],
\end{align}
and the equation for $f''$ can be obtained by taking the derivative of Eq.~\eqref{f'} with respect to $r$. Eqs.~\eqref{mu'}--\eqref{p'} will be used to write the tidal tensor in terms of the components of the metric and of the stress-energy tensor.

\subsection{Radial geodesics}\label{sec3}

In this section, we study the radial timelike geodesics in dirty BH spacetimes with line element \eqref{lineel} . We note that our formalism does not apply to the case of spinning particles, which are described by the Mathisson–Papapetrou–Dixon equations {\cite{2011CaQG...28.195025}. Radial timelike geodesics are such that $\dot{\theta}=\dot{\phi}=0$. The Lagrangian associated to the radial motion is~\cite{Wald::1984}~\footnote{We note that the Lagrangian for test particles in some alternative theories of gravity can differ from the one presented in Eq.~\eqref{lagrangian}~\cite{Pani:2011gy}. We are not going to consider these particular cases in this paper.}
\be
\label{lagrangian}\mathcal{L}=\frac{1}{2}g_{\mu\nu}\dot{x}^\mu\dot{x}^\nu=-\frac{1}{2},
\ee
where the overdots indicate derivative with respect to the proper time. From the Lagrangian~\eqref{lagrangian}, we obtain the following conserved quantities:
\begin{align}
	\label{energy_geo}&E=-\frac{\partial \mathcal{L}}{\partial \dot{t}}=f\,\dot{t},\\
	\label{Ang_mom}&L=\frac{\partial \mathcal{L}}{\partial \dot{\phi}}=r^2\,\sin^2\theta\,\dot{\phi}=0.
\end{align}
$E$ and $L$ are interpreted as the total energy and angular momentum of a test particle per unit mass, respectively. 
Since the particle is following a radial geodesic, we have $L=0$.
Substituting Eq.~\eqref{energy_geo} into Eq.~\eqref{lagrangian}, we find that the radial motion is described by
\be
\label{dotr}\dot{r}^2=\left(\frac{E^2}{f}-1\right)\,\left(1-\frac{2\,b}{r}\right).
\ee
Depending on the spacetime, if a test particle is released from rest at a radial coordinate $r_{i}$ outside the event horizon, with energy per unit mass equal to $E=\sqrt{f(r=r_{i})}$ [cf. Eq.~\eqref{dotr}], there is a point where the particle shall stop, at some radial coordinate $R_{\rm stop}$, and bounces back. The coordinate $R_{\rm stop}$ can be found using Eq.~(\ref{dotr}), through the equality $(E^2/f-1)=0$. 

In the next section, we investigate the geodesic deviation equation, and obtain its dependence with the energy $E$. The bouncing point given by $R_{\rm stop}$ should be taken into consideration to integrate the equation describing the geodesic deviation. We note, however, that for the dirty BHs explored here,  $f(r)$ is a monotonic function outside the event horizon. This implies that $R_{\rm stop}$ is always located inside the event horizon of the BH.
Being interested in the tidal forces in regions outside the event horizon, we will not explore the bounce radius in this paper.

\subsection{Tidal forces}\label{sec4}

Our main goal is to understand how the dirtiness of the BH spacetime influences the tidal forces. It is well-known that the relative acceleration between two infinitesimally nearby particles is described by the geodesic deviation equation \cite{Pirani,Manasse,dinverno,Hobson}:
\begin{align}\label{geodeviation}
	\frac{D^2\,\zeta^{\mu}}{D\tau^2}=K^{\mu}_{\ \ \nu}\,\zeta^{\nu},
\end{align}
where $\zeta^{\mu}$ is the infinitesimal displacement vector between two nearby geodesics, and  $K^{\mu}_{\ \ \nu}$ is the tidal tensor, given in terms of the Riemann tensor as
\be
K^{\mu}_{\ \ \nu}=R^{\mu}_{\ \ \alpha\,\beta\,\nu}\,u^\alpha\,u^\beta,
\ee
and $u^\mu$ is the unit vector tangent to the geodesic. 

 We can project the geodesic deviation vector and the tidal tensor components in an orthonormal tetrad basis. The projection is useful to analyze the tidal force felt by a body in the neighborhood of a 
  BH. We choose a tetrad basis attached to an observer in radial free fall in a dirty BH spacetime~\footnote{We can also study observers with non-zero angular momentum. In the case of circular motion, a particular choice of tetrads can be found in Ref.~\cite{2011AmJPhy...79..63}. Moreover, the effects of the angular momentum of a test particle on the tidal forces was recently studied in Ref.~\cite{Vandeev:2022}.}, which is given by~\cite{crispino2016}:
\begin{align}
	\label{tetrad0}&\lambda_{\hat{0}}\ ^\mu=\left(\frac{E}{f},\ -\left(1-\frac{2\,b}{r}\right)^\frac{1}{2}\sqrt{\frac{E^2}{f}-1},\ 0,\ 0\right),\\
	&\lambda_{\hat{1}}\ ^\mu=\left(-\frac{\sqrt{E^2-f}}{f},\ E\,f^{-\frac{1}{2}}\,\left(1-\frac{2\,b}{r}\right)^\frac{1}{2},\ 0,\ 0\right),\\
	&\lambda_{\hat{2}}\ ^\mu=\left(0,\ 0,\ r^{-1}, 0\right),\\
	\label{tetrad3}&\lambda_{\hat{3}}\ ^\mu=\left(0,\ 0,\ 0,\ r^{-1}\,\sin^{-1}\theta\right).
\end{align}
The indices with hat are tetrad basis indices. The vector $\lambda_{\hat{0}}\ ^\mu$ of the tetrad basis is timelike and it is equal to the four-velocity vector of the observer. The vectors ($\lambda_{\hat{1}}\ ^\mu$, $\lambda_{\hat{2}}\ ^\mu$, $\lambda_{\hat{3}}\ ^\mu$) are the three orthogonal spatial directions of the reference frame attached to the observer. The orthonormality condition of the tetrad basis is given by~\cite{chandrasekhar}:
\be
\lambda^\mu_{\hat{a}}\,\lambda^\nu_{\hat{b}}\,g_{\mu\,\nu}=\eta_{\hat{a}\,\hat{b}}, 
\ee 
where $\eta_{\hat{a}\,\hat{b}}=\text{diag}(-1\ ,\ 1,\ 1,\ 1)$ are the components of the Minkowski metric in Cartesian coordinates. The deviation vector $\eta^\mu$ and the tidal tensor $K^{\mu}_{\ \ \nu}$ can be expanded in terms of the orthonormal tetrad basis as \cite{chandrasekhar}
\begin{align}
	&\eta^\mu=\lambda^\mu_{\ \hat{a}}\,\eta^{\hat{a}},\\
	&K^\mu_{\ \ \nu}=\lambda^\mu_{\ \hat{a}}\,\lambda^{\ \hat{b}}_{\nu}\,K^{\hat{a}}_{\ \ \hat{b}},
\end{align}
respectively, where we have used the dual basis, given by
\be
\lambda^{\ \hat{b}}_{\mu}\equiv \lambda^\nu_{\ \hat{a}}\,g_{\mu\,\nu}\,\eta^{\hat{a}\,\hat{b}}.
\ee

We now impose that the metric functions obey the field equations \eqref{Einstein_eqs}. By computing the components of the Riemann tensor~\cite{Wald::1984} associated to the line element~\eqref{lineel}, and using Eqs.~\eqref{mu'}--\eqref{p'}, one can show that the tidal tensor, written in the tetrad basis attached to a radially infalling observer in a dirty BH spacetime, is given by
\begin{equation}
	{K^{\hat{a}}}_{\hat{b}}={\rm diag}(0,k_1,k_2,k_2),
\end{equation}
where
\begin{align}
	\label{k11}k_1& \equiv \frac{2 b (r)}{r^3}-4 \pi  (\rho+p-2 \sigma),\\
	\label{k22}k_2& \equiv -\frac{b (r)}{r^3}+4 \pi  \rho-\frac{4 \pi  E^2}{f} (\rho+p).
\end{align}

The tidal tensor components $k_1$ and $k_2$ for dirty BHs, given by Eqs.~\eqref{k11} and~\eqref{k22}, respectively, present interesting features. First, we point out that the well-known tidal forces in Schwarzschild spacetime can be obtained with $b(r)=M>0$, and $\rho=p=\sigma=0$. For the Schwarzschild case, the tidal force in the radial direction is always positive, since $k_1>0$. Besides that, the tidal forces in the angular directions are always negative, since $k_2<0$. However, due to the presence of the effective stress-energy tensor, the tidal forces in a dirty BH can differ from the corresponding Schwarzschild BH ones. This will depend on the equation of state, which relates the {energy density and pressures} quantities for a particular dirty BH. For instance, different tidal forces results can lead to different values for the Roche radius, so that a compact object would be tidally disrupted by a dirty BH in a different radial coordinate, when compared to the Schwarzschild BH case.

We also point out the dependence on $E^2$, present in Eq.~\eqref{k22}. This term is related to the difference in the strength of the angular tidal forces as measured in a radially infalling reference frame and in a static reference frame \cite{Horowitz:1997}. Therefore, an observer infalling radially in a dirty BH spacetime can measure different angular tidal forces, when compared to a static observer.

Interestingly, for the Schwarzschild solution, the term containing $E^2$ in Eq.~\eqref{k22} vanishes, since $\rho=p=0$. 
Moreover, for any dirty BH with \cite{Cho:2017nhx}
\begin{equation}\label{rhoep}
	\rho=-p,
\end{equation}
the dependence on $E^2$ is also absent, and hence the radially infalling observer and the static observer will agree on the measurement of the tidal forces. Equation~\eqref{rhoep} is satisfied for dirty BHs obeying \cite{lemos2018}:

\begin{equation}\label{lemos18}
	f(r)=h(r)=1-\dfrac{2\,b(r)}{r}.
\end{equation}
From Eq.~(\ref{k11}), if $\rho=-p$, the radial tidal force is determined only by $b$ and $\sigma$. Therefore, it may be possible to note some manifestations in the tidal forces due to the anisotropy in the stress-energy tensor. Eq.~\eqref{rhoep} also implies that Eq.~(\ref{k22}) is determined only by $b$ and $\rho$.

Let us now turn our attention to sign changes in the tidal forces, which happens, for instance, in the case of RN and Kerr BHs~\cite{crispino2016,Junior:2020yxg,R2,R5,R6}.
From Eqs.~\eqref{k11} and \eqref{k22}, we see that the tidal forces are zero when
\begin{align}
	\label{r0tf1}&\frac{2\,b (r_0^{\rm rtf})}{(r_0^{\rm rtf})^3}-4 \pi  \left(\rho(r_0^{\rm rtf})+p(r_0^{\rm rtf})-2 \sigma(r_0^{\rm rtf})\right)=0,\\
	\label{r0tf2}-&\frac{b (r_0^{\rm atf})}{(r_0^{\rm atf})^3}+4 \pi  \rho(r_0^{\rm atf})-\frac{4 \pi  E^2}{f(r_0^{\rm atf})}\left(\rho(r_0^{\rm atf})+p(r_0^{\rm atf})\right)=0,
\end{align}
are satisfied, where $r_0^{\rm rtf}$ and $r_0^{\rm atf}$ are the radial coordinates of the vanishing radial and angular tidal forces, respectively.

Moreover, using Eqs.~\eqref{horizon_cond1} and \eqref{horizon_cond2}, together with Eq.~\eqref{Gtt}, considering cases in which $\rho=-p$	, we find that at the event horizon the energy density satisfies
\be
\label{rho_horizon}8\pi r_+^2\rho(r_+)\leq 1,
\ee 
and from Eqs.~\eqref{horizon_cond1}, \eqref{k22} and \eqref{rho_horizon}, we obtain
\be
\label{k2_horizon}k_2(r_+)=\frac{1}{2r_+^2}\left(8\pi r_+^2\rho(r_+)-1\right)\leq 0.
\ee
Thus, dirty BHs present negative angular tidal forces at the event horizon. The equality in Eq.~\eqref{k2_horizon} holds for the extreme case, meaning that extreme dirty BHs always have null angular tidal forces at the event horizon.

In Sec.~\ref{sec:results}, we apply our results to some particular cases. We revisit some BH cases previously explored in the literature, but now considering them as dirty BHs, and we also analyze BH cases whose tidal forces have never been previously investigated. We focus on GR BH cases, but we emphasize that our analysis is also suitable to BHs in alternative theories of gravity. For instance, our general setup can be used to reproduce the tidal forces results in the context of holographic massive gravity, obtained in Ref.~\cite{Hong:2020}.



\section{Sample cases}\label{sec5}
\label{sec:results}
\subsection{Reissner-Nordstr{\"o}m black holes}

The spherically symmetric and electrically charged BH, also known as RN BH, can be regarded as a dirty BH, with the dirtiness being associated to the electromagnetic field~\cite{Visser1992}. In this case, we have
\begin{align}
	&f(r)=1-\frac{2\,M}{r}+\frac{Q^2}{r^2},\\
	&~\label{bRN}b(r)=M-\frac{Q^2}{2\,r},
\end{align}
and
\begin{equation}
	\rho= \frac{Q^2 }{8 \pi  r^4},~
	p=-\frac{Q^2}{8 \pi  r^4},~
	\sigma=-\frac{Q^2}{4 \pi  r^4},
\end{equation}
where $Q$ is the electric charge of the BH. The equations above can be translated in the following equations of state for an anisotropic fluid
\begin{equation}
	\rho=-p,~ \sigma=2p.
\end{equation}
Using Eqs.~\eqref{k11} and \eqref{k22}, we find the following tidal tensor components for RN BHs:
\begin{align}
	\label{k1RN}  &k_1=\frac{2M}{r^3}-\frac{3Q^2}{r^4},\\
	\label{k2RN}&k_2=-\frac{M}{r^3}+\frac{Q}{r^4},
\end{align}
which agree with the results found in Ref.~\cite{crispino2016}. We emphasize that for the RN spacetime we have $\rho=-p$, so that the tidal forces in the angular directions do not depend on $E^2$. The anisotropy contributes as $-2\,Q^{2}/r^{4}$ to the radial tidal force. The energy density contributes as $Q^{2}/2r^{4}$ to the angular tidal forces. In Fig.~\ref{fig:bsigmaRN}, we plot separately the contribution from the Misner-Sharp energy $b(r)$ and the contribution from the matter (which depends on $\sigma$ or $\rho$ solely) to the radial and angular components of the tidal force.
\begin{figure*}
	\centering\includegraphics[width=0.48\linewidth]{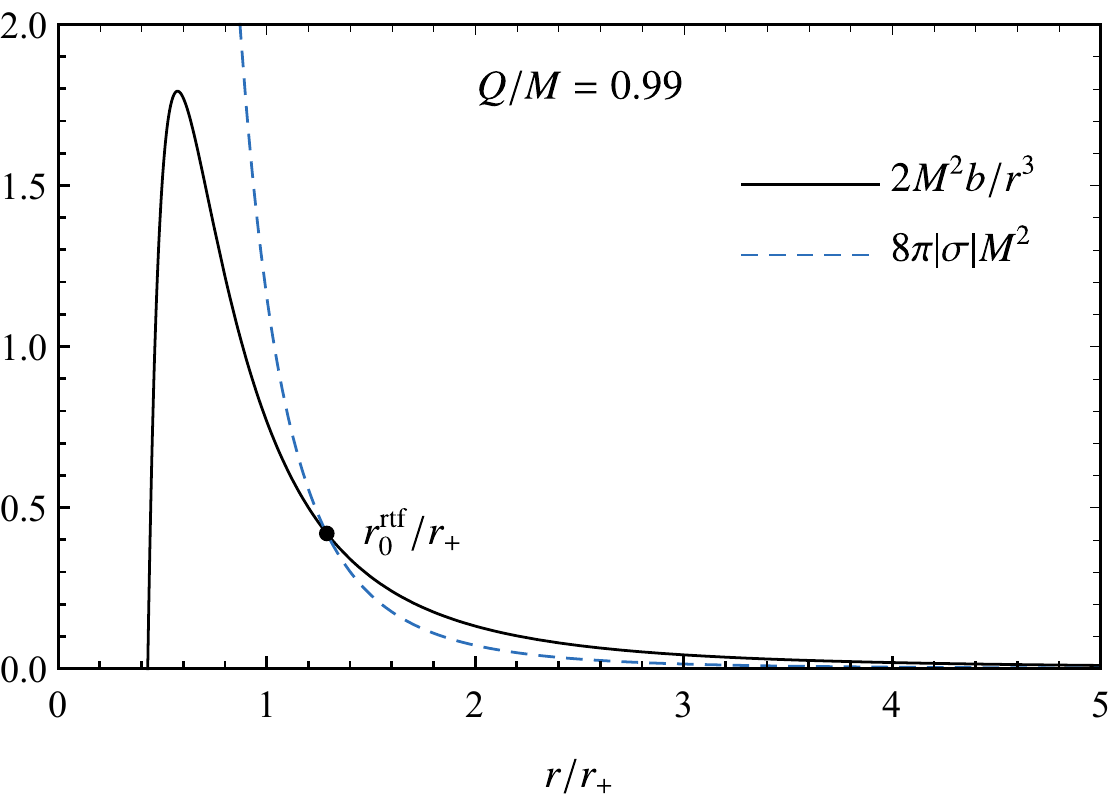}
	\centering\includegraphics[width=0.48\linewidth]{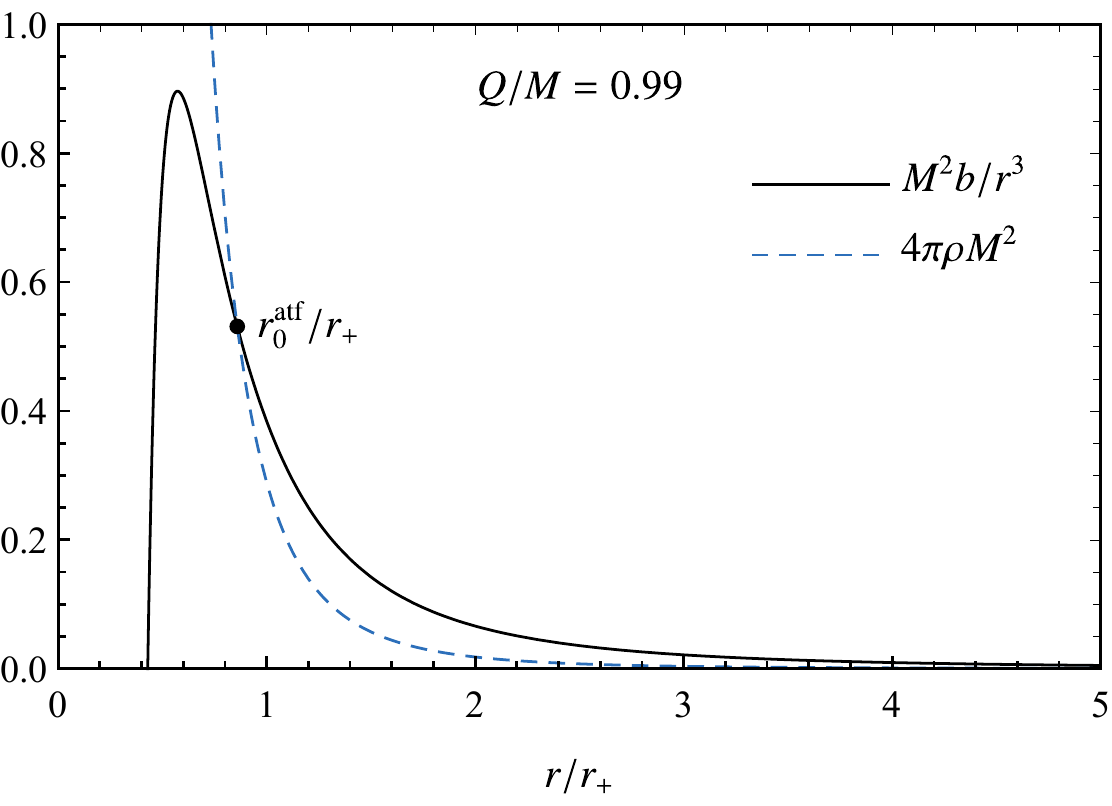}
	\caption{Contributions from the Misner-Sharp energy $b(r)$ (black solid lines) and from the matter (blue dashed lines) to the radial (left) and angular (right) tidal forces in a RN spacetime with $Q/M = 0.99$. The black disks denote the points in which the corresponding tidal force vanishes. Note that for this value of the BH charge, the radial tidal force vanishes outside the event horizon.}
	\label{fig:bsigmaRN}
\end{figure*} 
We can see that, for each component of the tidal force, there is a point in which these contributions have the same absolute value. This happens at the points $r_0^{\rm rtf}$ and $r_0^{\rm atf}$ in which  Eqs.~\eqref{r0tf1} and \eqref{r0tf2} are satisfied, respectively, implying that the components of the tidal force for radially infalling observers in RN spacetime vanish at (illustrated as a black disk in Fig.~\ref{fig:bsigmaRN})
\begin{align}
	&r_0^{\rm rtf}=\frac{3\,Q^2}{2\,M},\\
	&r_0^{\rm atf}=\frac{Q^2}{M}.
\end{align}
If $2\,\sqrt{2}/3\leq Q/M\leq 1$ the tidal force in the radial direction can vanish outside the event horizon, being in principle observable \cite{crispino2016}. This case shows the importance of the contribution from the anisotropy to the Eq.~(\ref{k1RN}), since the radial tidal force would not vanish outside the BH event horizon for isotropic dirtiness.

\subsection{Bardeen black holes}
\label{BardeenBH}
The Bardeen geometry is a regular BH solution, i.e., free of singularities. It was proposed by James M. Bardeen, illustrating that BHs are not necessarily singular \cite{bardeen1968}. Ayón-Beato and Garcia proposed that the Bardeen solution could be interpreted as a solution of Einstein's equations coupled to a nonlinear electrodynamics, described by the following Lagrangian~\cite{ayonbeato2000}:
\begin{align}
\mathcal{L}(F)=\frac{3}{2\,s\,g^2}\left(\frac{\sqrt{2\,g^2\,F}}{1+\sqrt{2\,g^2\,F}} \right)^\frac{5}{2},
\end{align}
where $s\equiv |g|/2M$, g is the magnetic monopole charge, $M$ is the mass, $F\equiv F_{\mu\nu}F^{\mu\nu}$ and $F_{\mu\nu}$ is the electromagnetic tensor.
Within our framework, the Bardeen solution can be regarded as a dirty BH, where the dirtiness comes from the nonlinear electromagnetic field.  In this case, we have
\begin{align}
	&f(r)=1-\frac{2\,M\,r^2}{\left(r^2+g^2\right)^\frac{3}{2}},\\
	&\label{bdebardeen}b(r)=\frac{M\,r^3}{\left(r^2+g^2\right)^\frac{3}{2}},
\end{align}
and
\begin{align}
	&\rho=\frac{3\,M\,g^2}{4\,\pi\,\left(r^2+g^2\right)^{\frac{5}{2}}},\\
	&p=-\frac{3\,M\,g^2}{4\,\pi\,\left(r^2+g^2\right)^{\frac{5}{2}}},\\
	&\sigma=-\frac{15\,M\,g^2\,r^2}{8\,\pi\,\left(r^2+g^2\right)^\frac{7}{2}},
\end{align}
The tidal tensor components for this spacetime are given by 
\begin{align}
	\label{k1BD}&k_{1}=\frac{2\, M}{\left(r^{2}+g^{2}\right)^{\frac{3}{2}}}\left(1-\frac{15\, g^{2}r^{2}}{2\left(r^{2}+g^{2}\right)^{2}}\right),\\
	\label{k2BD}&k_{2}=\frac{M}{\left(r^{2}+g^{2}\right)^{\frac{3}{2}}}\left(\frac{3\, g^{2}}{(r^{2}+g^{2})}-1\right),
\end{align}
which agree with the results found in Ref.~\cite{msharif2018}. We point out that $\rho=-p$, similarly to the RN case, so that the tidal forces as measured by a radially infalling or a static observer in Bardeen spacetime are the same. 

\begin{figure*}
	\centering
	\includegraphics[width=.48\linewidth]{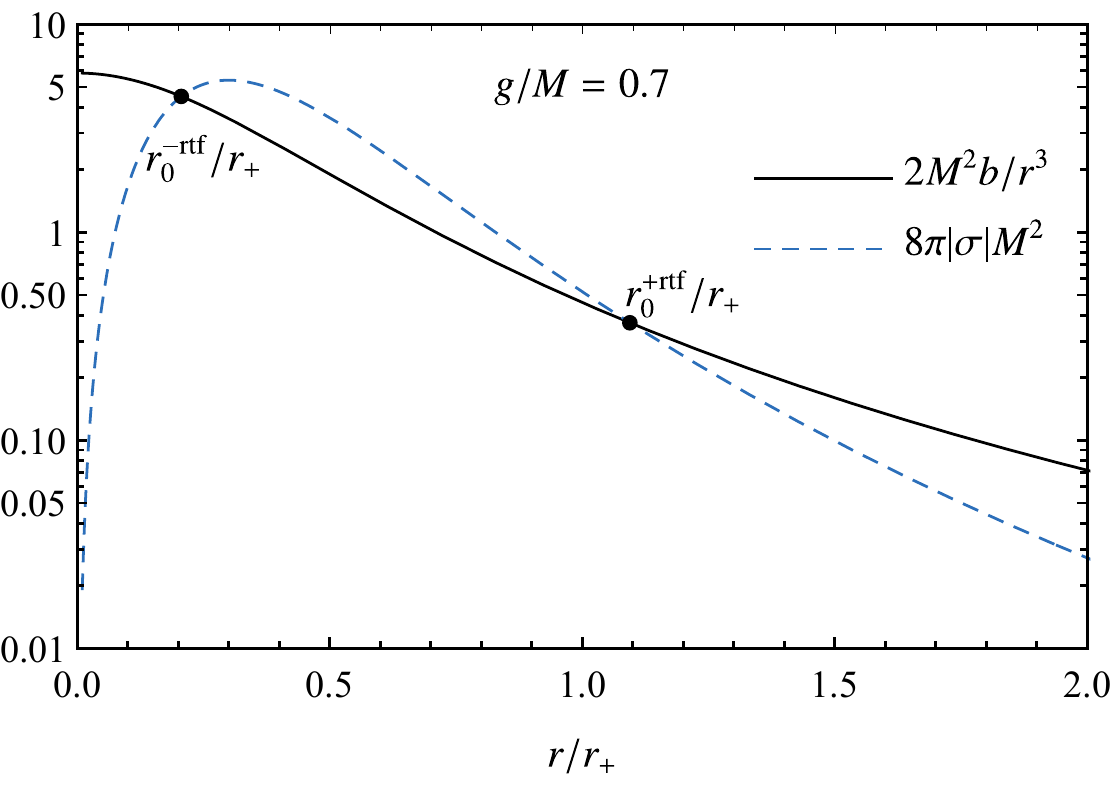}\includegraphics[width=.488\linewidth]{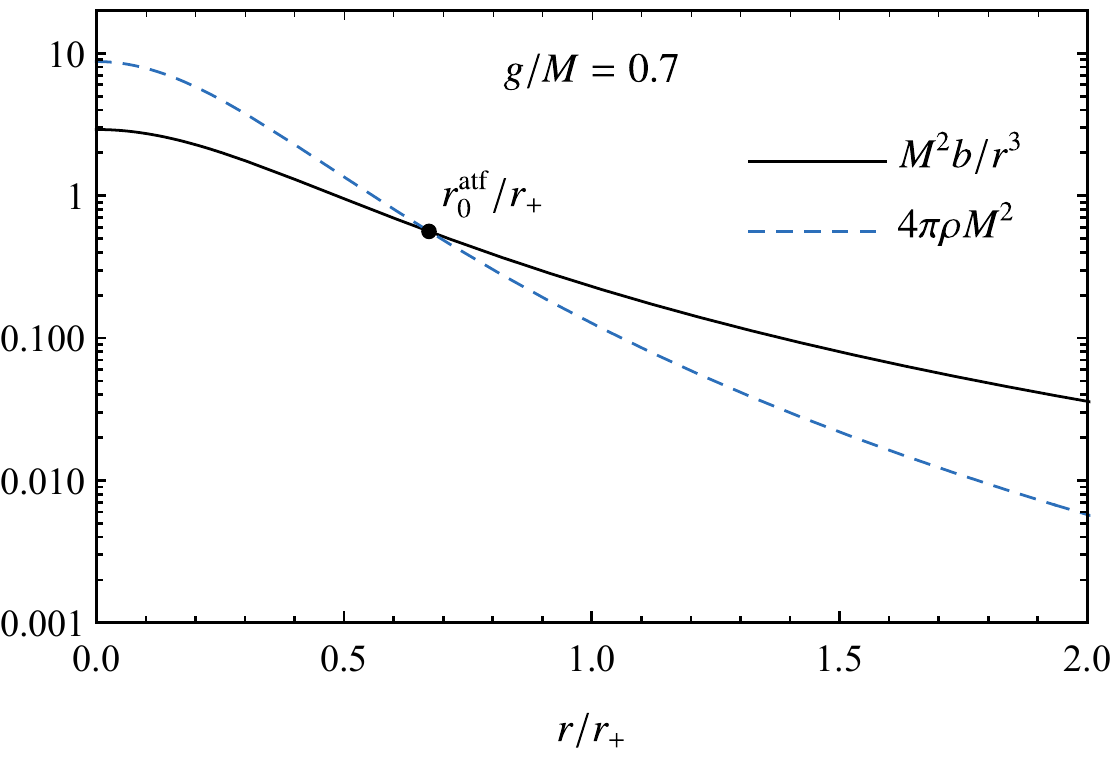}
	\caption{Contributions from the Misner-Sharp energy $b(r)$ (black solid lines) and from the matter (blue dashed lines) to the radial (left) and angular (right)  tidal forces for a Bardeen BH with $g/M = 0.7$.}
	\label{fig:bsigmabardeen}
\end{figure*}

The radial coordinates in which the radial and angular tidal forces vanish are given by 
\begin{align}
	\label{r0rtfBrdn}&r_0^{\pm \rm rtf}=\sqrt{11\pm\sqrt{105}}\,\frac{|g|}{2},\\
	&r_0^{\rm atf}=\sqrt{2}\,|g|,
\end{align} 
respectively.
Note that, in contrast to the RN case, the radial tidal force vanishes at two different locations.
If $r_0^{- \rm  rtf}<r<r_0^{+ \rm rtf}$, the anisotropy ($\sigma$) contribution (matter contribution) is dominant, hence the radial tidal force is negative, as we show in the left panel of Fig.~\ref{fig:bsigmabardeen}. In other words, in the region $r_0^{- \rm  rtf}<r<r_0^{+ \rm rtf}$ the body is compressed instead of being stretched in the radial direction. For $r\rightarrow 0$ the anisotropy goes to zero, so that the dirtiness is isotropic in this limit, in agreement with Ref.~\cite{marcos2018}. 

From the right panel of Fig.~\ref{fig:bsigmabardeen} we see that if $r<r_0^{\rm atf}$ the energy density ($\rho$) contribution (matter contribution) becomes greater than the contribution associated to $b(r)$, and the angular tidal forces are positive. Hence, the body is stretched in the angular directions for $r<r_0^{\rm atf}$. 

\begin{figure}[h!]
	\centering
	\includegraphics[width=1\linewidth]{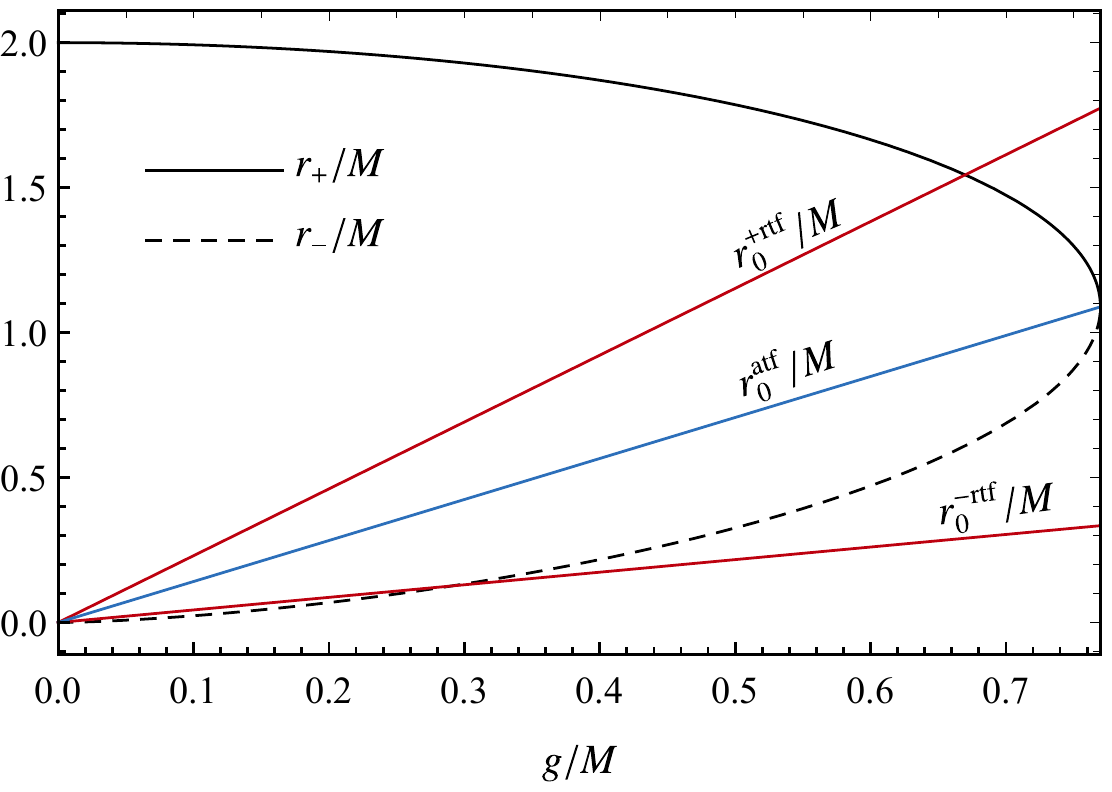}
	\caption{Radial coordinate of the event horizon ($r_{+}$), of the Cauchy horizon ($r_{-}$) and the radial coordinates where the radial ($r_{0}^{\pm rtf}$) and angular ($r_0^{atf}$) tidal forces vanish, plotted as functions of the monopole charge $g$.}
	\label{fig:FMNulaBardeen}
\end{figure}

In Fig.~\ref{fig:FMNulaBardeen}, we plot $r_{0}^{\pm \rm rtf}$, $r_0^{\rm atf}$, together with the radial coordinate of the Cauchy horizon ($r_{-}$) and of the event horizon ($r_{+}$). We note some similar features with the RN case (cf. Ref.~\cite{crispino2016}). For instance, the radial tidal force can vanish outside the event horizon. The angular tidal forces vanish between the event horizon and the Cauchy horizon, for any value of $g$. The radial tidal force vanishes in two different points of the radial coordinate, $r_{0}^{+ rtf}$ and $r_{0}^{- rtf}$, which are linear functions of $g$, in contrast to the RN case where it depends on the square of $Q$. For the extreme Bardeen BH case, the radial coordinates of the event horizon ($r_+$), of the Cauchy horizon ($r_-$) and of the vanishing angular tidal forces ($r_0^{atf}$) are the same.

\subsection{Generic regular magnetically charged black holes}
To obtain novel results for tidal forces in BH spacetimes, 
let us now turn our attention to a family of regular BHs associated with GR coupled to nonlinear electrodynamics. 
Spacetimes with a generic magnetically charged nonlinear electrodynamics source were introduced in Ref.~\cite{fanwang} and discussed in Refs.~\cite{bronnikov2017,Toshmatov2018}. Again (as in the Bardeen case), the dirtiness is associated to the nonlinear electromagnetic field. Considering only generic magnetically charged regular BH solutions, we have the following Lagrangian \cite{fanwang,Toshmatov2018}:
\begin{align}
&\mathcal{L}(F)=\frac{4\gamma}{\alpha}\frac{(\alpha\,F)^\frac{\nu+3}{4}}{\left[1+(\alpha\,F)^\frac{\nu}{4}\right]^{1+\frac{\gamma}{\nu}}},
\end{align}
and the spacetime geometry is described by
\begin{align}
	&\label{fGRBH}f(r)=1-\frac{2\,M\,r^{\gamma-1}}{\left(r^{\nu}+g^{\nu}\right)^{\frac{\gamma}{\nu}}},\\
	&b(r)=\frac{M\,r^{\gamma}}{\left(r^{\nu}+g^{\nu}\right)^{\frac{\gamma}{\nu}}},
\end{align}
and
\begin{align}
	&\rho(r)=\frac{\gamma\,M\,g^{\nu}\,r^{\gamma-3}}{4\,\pi\,\left(r^{\nu}+g^{\nu}\right)^{\frac{\gamma}{\nu}+1}},\\
	&p(r)=-\frac{\gamma\,M\,g^{\nu}\,r^{\gamma-3}}{4\,\pi\,\left(r^{\nu}+g^{\nu}\right)^{\frac{\gamma}{\nu}+1}},\\
	&\sigma(r)=\frac{\gamma\,M\,g^{\nu}\,r^{\gamma-3}\,(g^{\nu}(\gamma-3)-r^{\nu}(3+\nu))}{8\,\pi\,\left(r^{\nu}+g^{\nu}\right)^{2+\frac{\gamma}{\nu}}},
\end{align}
where $g$ is the magnetic charge parameter of the BH, $\gamma\ge 3$ is related to the electrodynamics non-linearity, $M=  g^{3} / \alpha$ is the mass of the BH (and is equal to the electromagnetically induced mass \cite{Toshmatov2018}), with $\alpha$ being an intensity parameter related to the charge, and $\nu>0$ is a dimensionless constant. The tidal tensor components for this spacetime are given by
\begin{align}
	\label{k1GB}
	&k_{1}=\frac{2\,M\,r^{\gamma-3}}{\left(r^{\nu}+g^{\nu}\right)^{\frac{\gamma}{\nu}}}\left[1+\frac{\gamma\,g^{\nu}\left(g^{\nu}(\gamma-3)-r^{\nu}(3+\nu)\right)}{2 \left(r^{\nu}+g^{\nu}\right)^{2}}\right],\\
	\label{k2GN}&k_{2}=\frac{M\,r^{\gamma-3}}{\left(r^{\nu}+g^{\nu}\right)^{\frac{\gamma}{\nu}}}\left[\frac{\gamma\,g^{\nu}}{\left(r^{\nu}+g^{\nu}\right)}-1\right].
\end{align}
We point out that, for this generic regular magnetically charged BH spacetime, we also have $\rho=-p$ (as for the RN and Bardeen BH cases), so that the tidal forces for a radially infalling and a static observer are the same. The second term between the squared brackets in Eq.~(\ref{k1GB}) is the contribution due the anisotropy ($\sigma$), and the first term between the squared brackets in Eq.~(\ref{k2GN}) is due to the energy density ($\rho$). 
As particular cases, we have the Bardeen BHs, for $\gamma=3$ and $\nu=2$; and the Hayward BHs, for $\gamma=3$ and $\nu=3$.

The tidal forces in this generic regular magnetically charged BH spacetime vanish at
\begin{align}\label{r0rtfgeneric}
	&r_0^{\pm \rm  rtf}=\frac{|g|}{4^{\frac{1}{\nu}}}\left[-4+\gamma\left(3+\nu\right)\pm\sqrt{\gamma}\sqrt{-8\,\nu+\gamma\left(1+\nu\left(6+\nu\right)\right)}\right]^{\frac{1}{\nu}},\\
	\label{ratf}&r_0^{\rm atf}=|g|\left(\gamma-1\right)^{\frac{1}{\nu}}.
\end{align}
If $\gamma \neq 3$, the tidal tensor also vanishes at
\begin{equation}
	r_{0}=r_{0}^{\rm rtf}=r_{0}^{\rm atf}=0.
\end{equation} 
We point out that the tidal forces in Bardeen BHs do not vanish at $r=0$ (see Sec.~\ref{BardeenBH}), since $\gamma=3$.

Next, let us analyze regular BHs with $\nu=2$, other than the Bardeen BH, which are called Bardeen-like BHs \cite{Toshmatov2018}. As an illustrative example, we consider the $\gamma=4$ case [for the Bardeen-like BHs ($\nu=2$)]. The contribution related to the Misner-Sharp energy (which depend on $b$) and the one related to the properties of the dirtiness (matter contribution) can be obtained by substituting the values of $\gamma$ and $\nu$ in the equations (\ref{fGRBH})-(\ref{ratf}). 

In Fig.~\ref{fig:rhoebmu4nu2}, we plot the $k_2$ component of the tidal tensor (inset), as well as the separated contributions from the energy density $\rho (r)$ (dirtiness term) and from the $b(r)$ function. There are two points for which the angular tidal forces are equal to zero, namely $r_{0}^{\rm atf}$ and $r=0$. The first is due to the cancellation of the contributions from the dirtiness and the  $b(r)$ term, while the latter occurs where the contributions from the dirtiness and the $b(r)$ terms vanish.  

\begin{figure}[h!]
	\centering
	\includegraphics[width=1\linewidth]{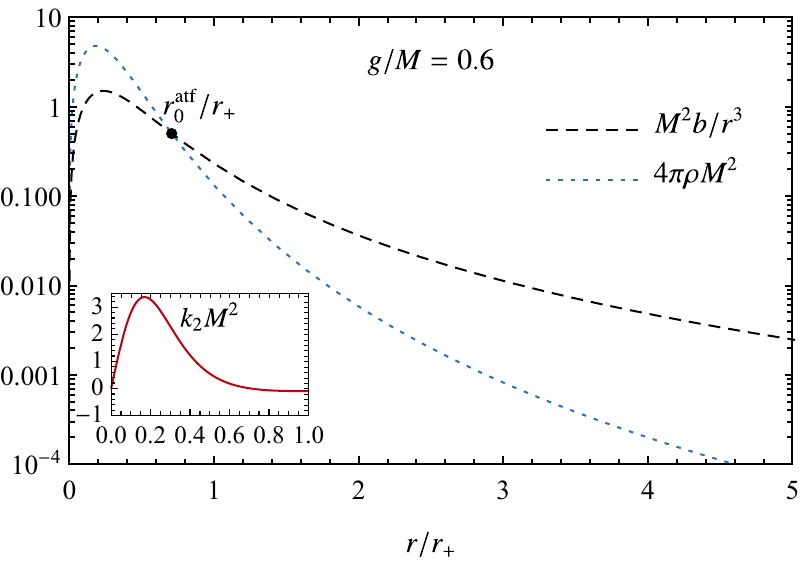}
	\caption{Tidal tensor component $k_2$, plotted in the inset, and the separated contributions to it, one from the energy density $\rho$ (blue dotted line, dirtiness term) and the other from the Misner-Sharp energy $b(r)$ (black dashed line) to the angular tidal forces, for Bardeen-like BHs ($\nu=2$) with $\gamma=4$.} 
	\label{fig:rhoebmu4nu2}
\end{figure}

In Fig.~\ref{fig:bsigmamu4nu2}, we plot the $k_1$ component of the tidal tensor (inset), and also the separated contributions from the anisotropy $\sigma$ (dirtiness term) and the $b(r)$ function  (Misner-Sharp energy). There are three points for which the radial tidal force vanishes. For $r\rightarrow 0$, the radial tidal force vanishes because the contributions from the anisotropy $\sigma (r)$ (dirtiness term) and from the $b(r)$ function (Misner-Sharp energy) vanish at the center of the BH.

\begin{figure}
	\centering
	\includegraphics[width=1\linewidth]{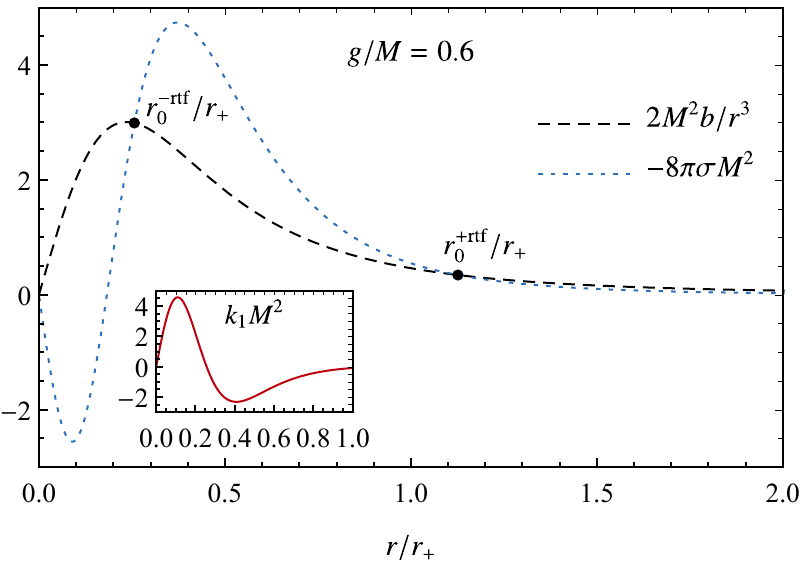}
	\caption{Tidal tensor component $k_1$, plotted in the inset, and the separated contributions to it, one from the anisotropy $\sigma$ (blue dotted line, dirtiness term) and the other from the Misner-Sharp energy $b(r)$ (black dashed line) to the radial tidal force, for Bardeen-like BHs ($\nu=2$) with $\gamma=4$.}
	\label{fig:bsigmamu4nu2}
\end{figure}

In Fig.~\ref{fig:FMNulamu4nu2}, we show the radial coordinate of the horizons, as well as the values of the radial coordinate where the tidal forces vanish.  Some features present in the Bardeen BH case are also manifest here. For instance, the linear dependence of $r_{0}^{\pm rtf}$ with the monopole charge, the vanishing of the angular tidal forces between the horizons, and the vanishing of the radial tidal force outside the event horizon, for $g\,>\,0.567\,M$. Moreover, for the extreme case, the radial coordinate of the event horizon, of the Cauchy horizon and of the vanishing angular tidal forces are the same.

\begin{figure}
	\centering
	\includegraphics[width=1\linewidth]{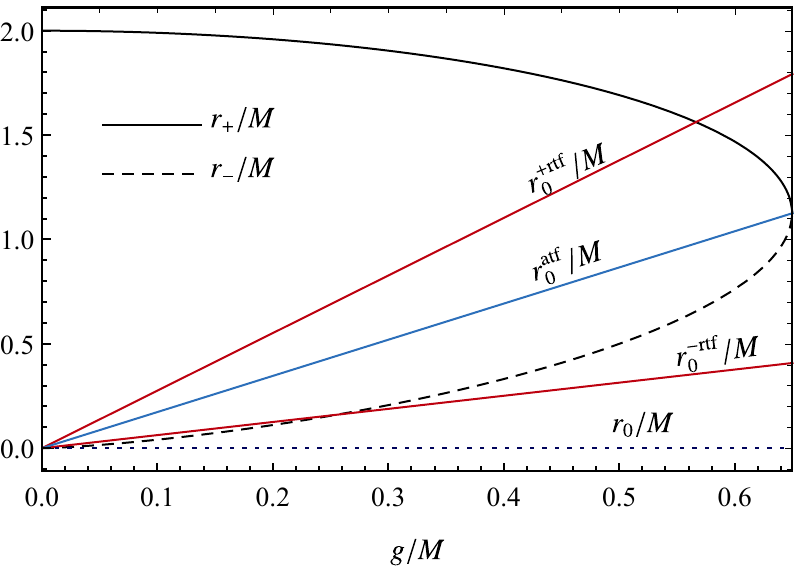}
	\caption{Radial coordinates of the horizons ($r_{\pm}$) and the radial coordinates where the tidal forces vanish ($r_{0}^{\pm rtf}, r_0^{atf}$), as functions of the monopole charge $g$, for Bardeen-like BHs ($\nu=2$) with $\gamma=4$.}
	\label{fig:FMNulamu4nu2}
\end{figure}

\subsection{Schwarzschild black hole with a surrounding thin spherical shell}
We consider the tidal forces in a Schwarzschild BH surrounded by a thin spherical shell~\cite{Caio2016,Israel,Macedo:2018chs}. The shell is regarded as constituted by a perfect fluid and satisfies the Tolman-Oppenheimer-Volkoff (TOV) equations \cite{DBH,Caio2016}. In this case, the metric functions are given by
\begin{align}
	f(r)&=\begin{cases}
		\beta(1-\frac{2\,M_{\rm BH}}{r}), & r < R_{S}, \\
		(1-\frac{2\,M}{r}), \, &r > R_{S},
	\end{cases} \\
	b(r)&=\begin{cases}
		M_{\rm BH}, &r < R_{S}, \\
		M, &r > R_{S},
	\end{cases}
\end{align}
where $\beta$ is a constant such that $f(r)$ is continuous across the shell, $M$ is the Arnowitt-Deser-Misner (ADM) mass, and $M_{\rm BH}$ is the mass associated to the area of the BH event horizon. The tidal forces for a freely falling observer have essentially the same algebraic form as in the Schwarzschild case, the difference being in the nature of the mass term:
outside the shell it is the ADM mass $M$,  while inside the shell it is the mass associated to the area of the event horizon $M_{\rm BH}$, namely
\begin{align}
	k_{1}&=
	\begin{cases}
		\dfrac{2M}{r^{3}},& r>R_S,\\
		\dfrac{2\,M_{\rm BH}}{r^{3}},& r<R_S,
	\end{cases}\\
	k_{2}&=
	\begin{cases}
		-\dfrac{M}{r^{3}},& r>R_S,\\
		-\dfrac{M_{\rm BH}}{r^{3}},& r<R_S.
	\end{cases}
\end{align}

\begin{figure*}
	\centering\includegraphics[width=0.48\linewidth]{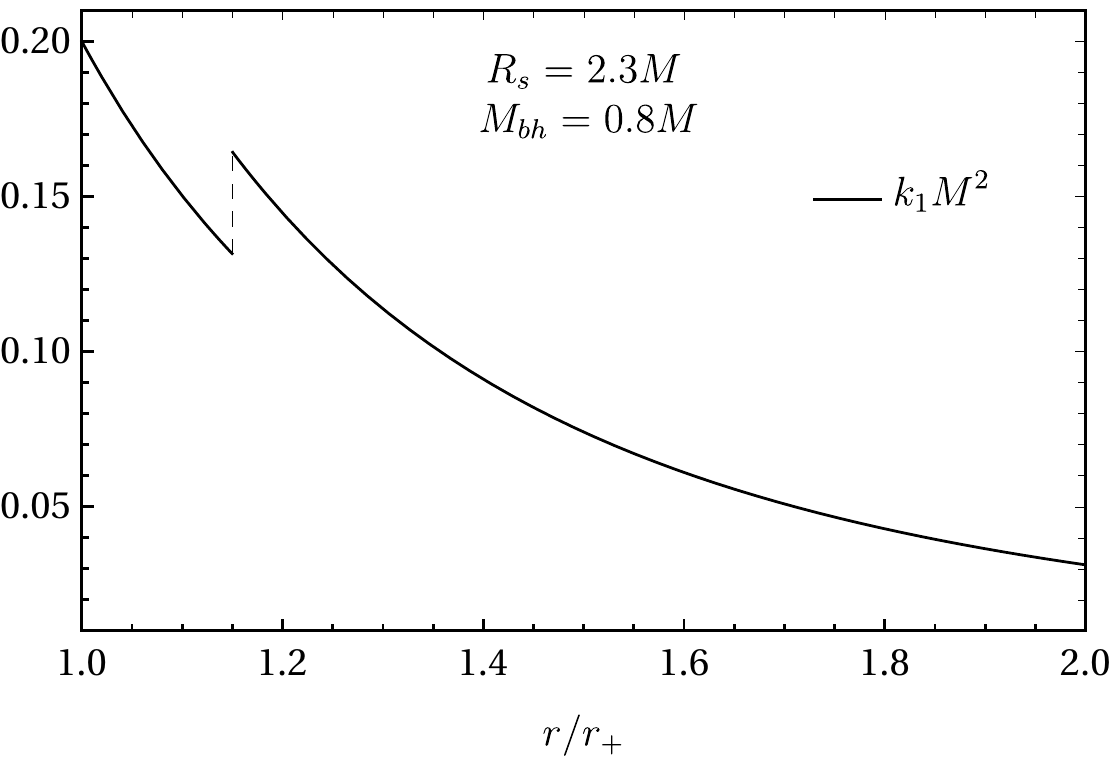}
	\centering\includegraphics[width=0.488\linewidth]{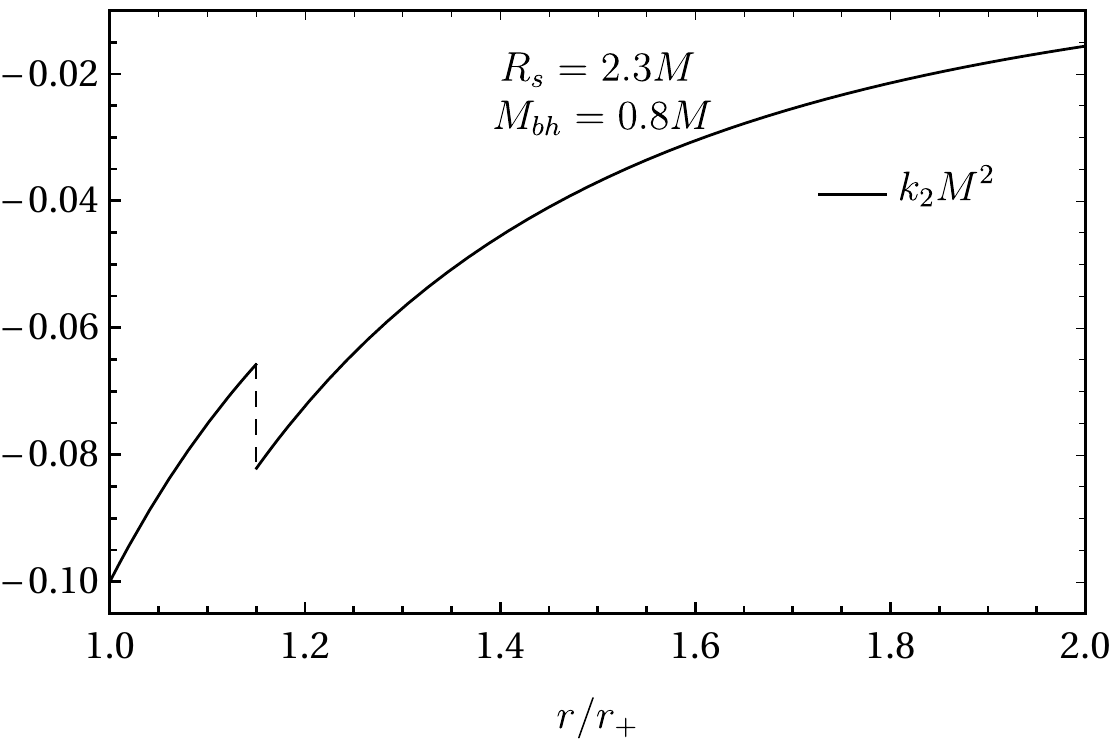}
	\caption{Tidal tensor components $k_1$ (left) and $k_2$ (right) in a Schwarzschild BH surrounded by a thin spherical shell for $R_s=2.3M$ and $M_{BH}=0.8M$.}
	\label{fig:tidalforce_Schdbh}
\end{figure*} 

In this case the dirtiness is localized in the thin spherical shell, allowing us to clearly illustrate an interesting effect of matter surrounding the BH: 
Due to the separation of the total mass of the system, now composed by the BH and the spherical shell,
the tidal forces in the region between the BH horizon and the position of the spherical shell are lower (in modulus) than the tidal forces of an isolated Schwarzschild BH with the same ADM mass. In Fig.~\ref{fig:tidalforce_Schdbh}, we show the components $k_1$ and $k_2$ of the tidal tensor for a Schwarzschild BH surrounded by a spherical shell for $R_s=2.3M$ and $M_{BH}=0.8M$, where we note the discontinuity on the tidal forces at the spherical shell.
We can quantify how smaller the tidal forces are, compared with the isolated Schwarzschild BH case, by directly comparing $M_{\rm BH}$ and $M$, through the ratio $M_{\rm BH}/M<1$. The present tidal forces analysis can be generalized to the case of $N-$spherical shells~\cite{Israel}.

\subsection{Black holes surrounded by anisotropic matter}
As a final example, we apply our procedure to analyze the tidal forces in a newly presented BH solution, obtained by considering anisotropic matter minimally coupled to gravity~\cite{Cardoso:2021wlq}. For this particular case, the radial pressure is identically zero. The matter terms are given by
\begin{equation}
	\rho=\frac{M(a_0+2M_{\rm BH})(1-2M_{\rm BH}/r)}{2\pi r (r+a_0)^3},~q=\frac{b(r)\rho}{2(r-2b(r))},
\end{equation}
where $M$ is related to the mass of the matter, $M_{\rm BH}$ the mass associated to the BH horizon, and $a_0$ is a length scale related to the distribution of matter ($a_0\gtrsim 10^4M$ for galaxies distributions~\cite{Cardoso:2021wlq}). The Misner-Sharp energy $b(r)$ is given by
\begin{equation}
	b(r)=M_{\rm BH}+\frac{M r^2}{(a_0+r)^2}\left(1-\frac{2M_{\rm BH}}{r}\right)^2,
\end{equation}
from which it is clear that the BH horizon is located at $r_{\rm h}=2M_{\rm BH}$ and that the total mass is given by $M+M_{\rm BH}$. Finally, the metric function is given by	
\begin{equation}
	f(r)=\left(1-\frac{2M_{\rm BH}}{r}\right)e^{\gamma},
\end{equation}
where
\begin{align}
	\gamma&=-\pi\sqrt{\frac{M}{\varepsilon}}+2\sqrt{\frac{M}{\varepsilon}}\arctan\left(\frac{r+a_0-M}{\sqrt{M\varepsilon}}\right),\\
	\varepsilon&=2a_0-M+4M_{\rm BH}.
\end{align}
With the above quantities, we can compute the tidal tensor components $(k_1,k_2)$ and compare them to the isolated Schwarzschild BH case. Due to the lengthy form of the functions, we shall not display them explicitly here, limiting ourselves to show some particular cases.

\begin{figure*}
	\centering\includegraphics[width=0.48\linewidth]{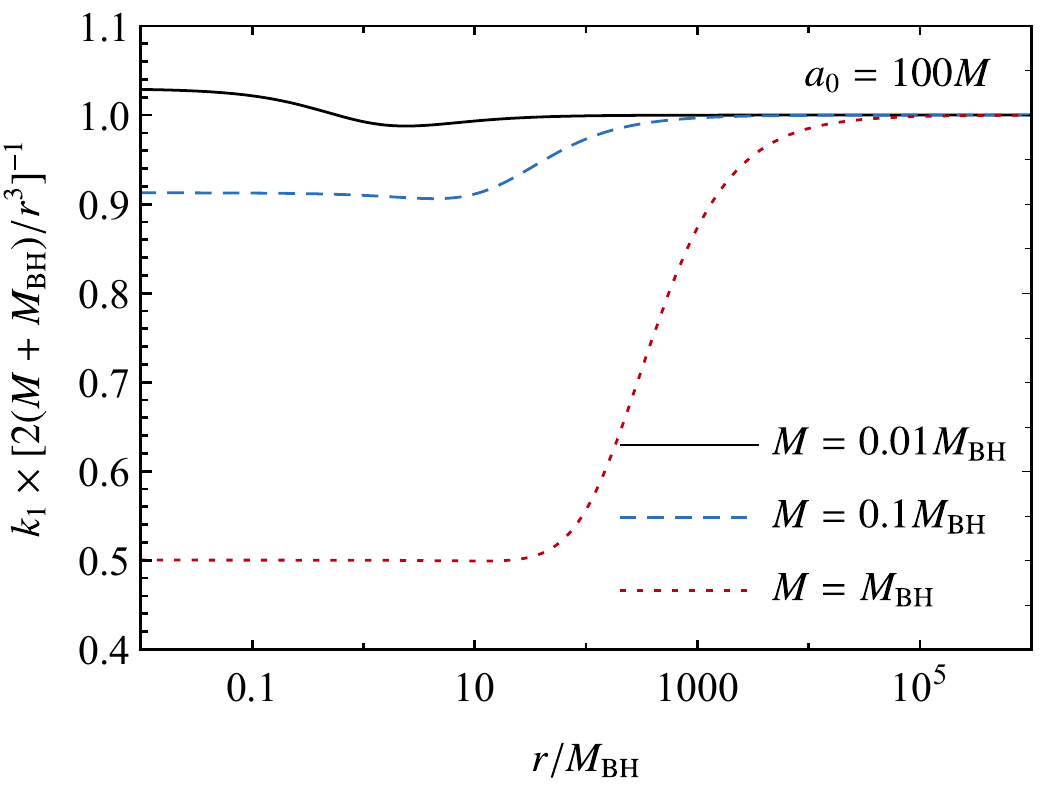}\includegraphics[width=0.48\linewidth]{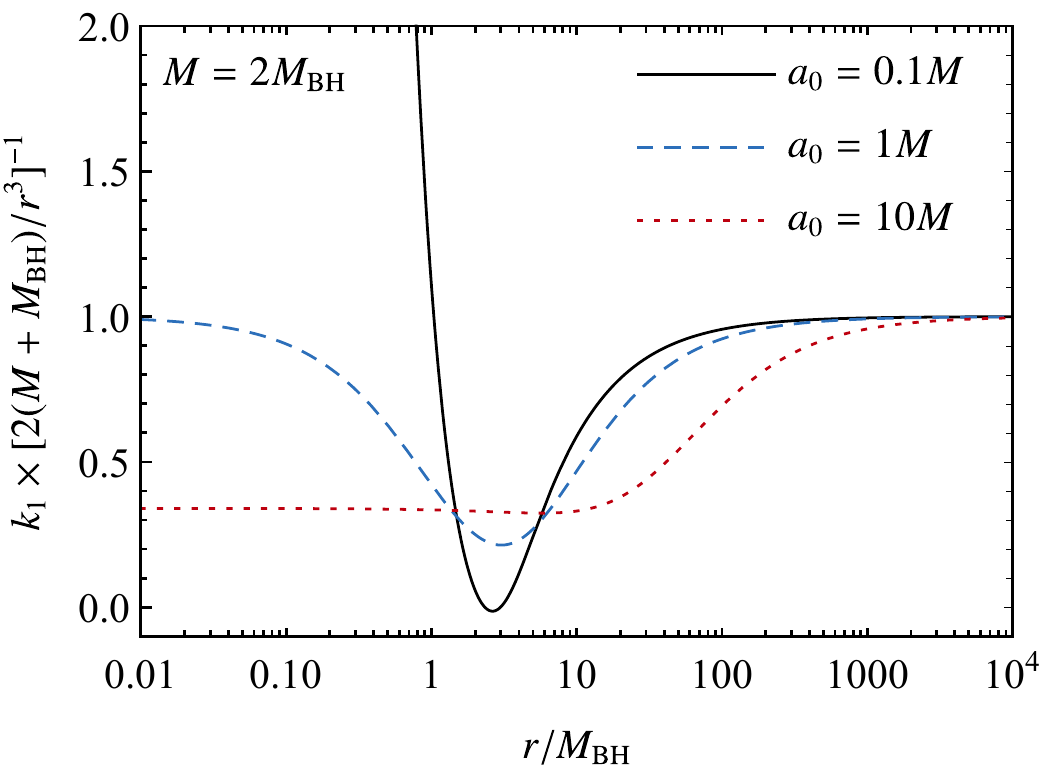}
	\caption{
		Normalized tidal tensor component $k_1$, for BHs with anisotropic matter, as a function of the radius, for some representative cases. In the left panel we fix $a_0=100M$, considering different values for the mass $M$. In the right panel we fix $M=2M_{\rm BH}$, considering different values for the length scale $a_0$. In both cases, we observe considerable deviations from the Schwarzschild BH case.}\label{fig:aniso1}
\end{figure*}
In Fig. \ref{fig:aniso1} we show the radial tidal tensor component $k_1$ for the BH surrounded by anisotropic matter, normalized by the Schwarzschild case. The dirty BH is asymptotically flat, so that for large distances the ratio tends to the unity. In the left panel of Fig.~\ref{fig:aniso1} we fix the value for $a_0=100M$, varying $M$. We can see that even for the case in which the mass $M$ represents $10\%$ of the total mass, the result can deviate notably from the Schwarzschild case, mainly near the BH (notice that $M=0$ represents the Schwarzschild case). In the right panel of Fig.~\ref{fig:aniso1} we turn our attention to changes in $a_0$, fixing $M$. We focus on some specific cases to show that, within the parameter space, there is a possibility of having repulsive tidal forces considering anisotropic matter (notice the black solid curve).

\begin{figure*}
	\centering\includegraphics[width=0.48\linewidth]{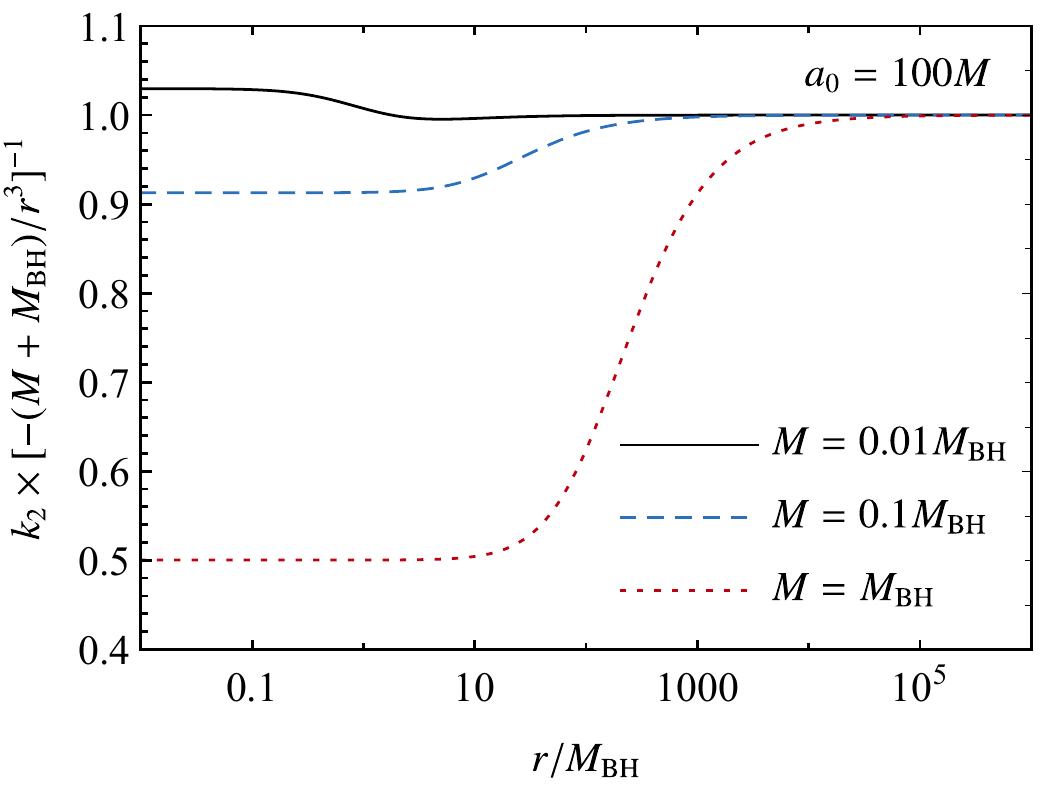}\includegraphics[width=0.48\linewidth]{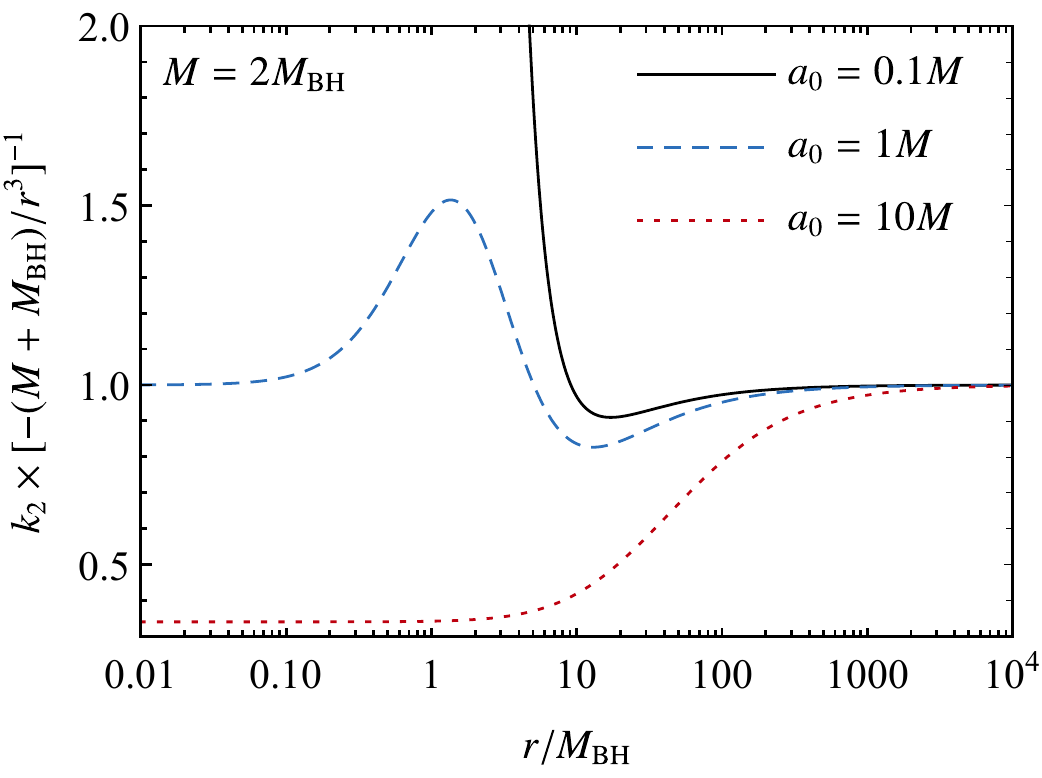}
	\caption{
		Normalized tidal tensor component $k_2$, for BHs with anisotropic matter, as a function of the radius, for some representative cases. We use the same parameter space as the one described in Fig.~\ref{fig:aniso1}, considering, in addition, $E=1$. We see that the deviation from Schwarzschild case is considerable. For a fixed value of $a_0$ (left panel) the modifications are similar to the ones observed in Fig.~\ref{fig:aniso1}.}\label{fig:aniso2}
\end{figure*}
For the angular component of the tidal tensor we have to fix a value for the specif energy $E$. For simplicity, we shall focus on the $E=1$ case, investigating the same parameter space of Fig.~\ref{fig:aniso1}. The result is shown in Fig.~\ref{fig:aniso2}. We see, for instance, that for $a_0=100M$ (left panel), the changes in the normalized angular tidal tensor are very similar to the radial one. 
Considering a fixed value of $M$ and changing $a_0$, however, we see that the behavior is distinctive (right panel) and we do not observe changes in the sign of the tidal tensor component $k_2$. We conclude that for these configurations, the angular tidal force is always attractive.

\section{Final remarks}\label{sec6}
We studied the tidal tensor for generic spherically symmetric dirty BH spacetimes, and investigated the influence of the dirtiness on the tidal forces. An important feature is the influence of the energy ($E$) of the particle in the angular tidal forces equations. For dirty BHs with  $\rho=-p$, the energy ($E$) does not contribute to the angular tidal forces, as shown explicitly in the examples we analyzed. Moreover, we have shown that the angular tidal forces at the event horizon are always less or equal to zero, and the equality holds only for extreme dirty BHs. We also noted that only the anisotropy ($\sigma = p-q$) contributes to the radial tidal force, while for the angular tidal forces only the energy density ($\rho$) contributes. 
We focused on some particular cases, describing (electrically or magnetically) charged BHs, and also a Schwarzschild BH surrounded by a spherical shell. 

For the RN BH, the function $b(r)$ is positive outside the event horizon, so that the negative contribution for the radial tidal force comes only from the anisotropy in the stress-energy tensor associated to the electromagnetic field. On the other hand, the energy density contributes positively to the angular components of the tidal forces, and tidal forces in the angular directions changes sign between the event horizon and the Cauchy horizon. The RN BH has a singularity at $r=0$, hence the tidal forces grow indefinitely as the singularity is approached. However, a radially infalling observer bounces back before reaching the singularity \cite{crispino2016}.

For the Bardeen BH, the function $b(r)$ is positive outside the event horizon, so that the negative contribution for the radial tidal force comes only from the anisotropy ($\sigma$), and it is responsible for the change from the stretching to the compression of the body, similarly to the RN BH case. The radial tidal force vanishes in two points, $r_{0}^{\pm rtf}$, which depend linearly on the magnetic  charge. The angular tidal forces change sign between the event horizon and the Cauchy horizon. The occurrence of a point of vanishing angular tidal forces, $r_{0}^{atf}$, is a consequence of the positive contribution from the energy density.

For the case of more general magnetically charged regular BHs determined by additional parameters $\nu$ and $\gamma$, the contribution from the anisotropy for the radial tidal force can be positive [although this is not the case of the Bardeen BH, which has ($\nu = 2$ and) $\gamma=3$]. 
We analyzed the tidal forces in a Bardeen-like BH ($\nu=2$) with $\gamma=4$. In this case: (i) the radial tidal force vanishes at three values of the radial coordinate, namely at $r_{0}^{\pm rtf}$ and at the center of the BH; and (ii) the angular tidal forces vanish at two points, namely at $r_{0}^{atf}$ and also at the center of the BH.

We investigated the case of a Schwarzschild BH with a thin spherical shell. 
Although the expressions of the tidal forces are algebraically the same as in the isolated Schwarzschild BH case, 
there is a discontinuity in the $b(r)$ function. For such configuration, in the region between the event horizon and the spherical shell, the tidal forces are lower (in modulus) than the tidal forces of an isolated Schwarzschild BH with the same ADM mass.
This is a consequence of the fact that the total mass of the system is diluted around the central BH rather than concentrated as an isolated Schwarzschild BH.

To further illustrate the effects of the matter terms, we  investigated the case in which the BH is surrounded by a specific anisotropic distribution of matter. This solution was recently presented in Ref.~\cite{Cardoso:2021wlq}. We have shown that, even when the matter contribution is small, the deviations from the tidal tensor components of the Schwarzschild spacetime can be non-negligible. We have also shown that the radial component of the tidal tensor can vanish in a position located outside the event horizon.

We note that the description presented here is general, in the sense that it can be extended to deal with other astrophysically relevant setups. For instance, it is argued that dark matter can accumulate in the surrounding of compact stars and, therefore, making them dirty~\cite{Brito:2015yfh}.  We can straightforwardly include the quantities describing the dark matter components in the matter functions of equations \eqref{k11} and \eqref{k22}, obtaining their contributions to the tidal force. 
Additionally, the analysis presented in this paper can be naturally extended to investigate tidal forces modifications introduced by general accretion disks surrounding BHs.

\begin{acknowledgements}
The authors would like to thank Funda\c{c}\~ao Amaz\^onia de Amparo a Estudos e Pesquisas (FAPESPA), Conselho Nacional de Desenvolvimento Científico e Tecnológico (CNPq) and Coordenação de Aperfeiçoamento de Pessoal de Nível Superior (CAPES) -- Finance Code 001, from Brazil, for partial financial support. This work has also been supported by the European Union's Horizon 2020 research and innovation (RISE) program H2020-MSCA-RISE-2017 Grant No. FunFiCO-777740.

\end{acknowledgements}

\end{document}